\documentclass[prd, twocolumn, nofootinbib, floatfix]{revtex4-1}

\usepackage{amsmath}
\usepackage{graphicx}
\usepackage{dcolumn}
\usepackage{bm}
\usepackage{enumerate} 
\usepackage{epsfig}
\usepackage{amssymb,latexsym,mathrsfs}
\usepackage{graphicx}
\usepackage{color}
\usepackage{hyperref}

\hypersetup{
    colorlinks=true,
    linkcolor=red,
    citecolor=blue,
} 

\newcommand{\be}{\begin{equation}}
\newcommand{\ee}{\end{equation}}
\newcommand{\beq}{\begin{equation}}
\newcommand{\eeq}{\end{equation}}

\newcommand{\bs}{\begin{split}} 
\newcommand{\bea}{\begin{eqnarray}}
\newcommand{\eea}{\end{eqnarray}}
\newcommand{\beqa}{\begin{eqnarray}}
\newcommand{\eeqa}{\end{eqnarray}}

\newcommand{\Om}{\Omega_m}
\newcommand{\omhh}{\Omega_m h^2}

\newcommand{\tfit}{\widetilde{\Delta t}} 
\newcommand{\ttrue}{\Delta t} 
\newcommand{\tdd}{D_{\Delta t}}

\begin{document}

\title{Next Generation Strong Lensing Time Delay Estimation with 
Gaussian Processes} 
\author{Alireza Hojjati$^{1,2}$ \& Eric V.\ Linder$^3$} 
\affiliation{$^1$Dept.~of Physics and Astronomy, University of British
Columbia, Vancouver, BC V6T 1Z1, Canada\\ 
$^2$ Physics Department, Simon Fraser University, Burnaby, BC V5A 1S6, Canada\\
$^3$Berkeley Center for Cosmological Physics \& Berkeley Lab, 
University of California, Berkeley, CA 94720, USA} 

\begin{abstract}
Strong gravitational lensing forms multiple, time delayed images of 
cosmological sources, with the ``focal length'' of the lens serving as a 
cosmological distance probe. Robust estimation of the time delay distance 
can tightly constrain the Hubble constant as well as the matter density and 
dark energy. Current and next generation surveys will find hundreds to 
thousands of lensed systems but accurate time delay estimation from noisy, 
gappy lightcurves is potentially a limiting systematic. Using a large sample of 
blinded lightcurves from the Strong Lens Time Delay Challenge we develop and 
demonstrate a Gaussian Process crosscorrelation technique that delivers an 
average bias within 0.1\% depending on the sampling, necessary for 
subpercent Hubble constant determination. The fits are accurate (80\% of 
them within 1 day) for delays from 5--100 days and robust against cadence 
variations shorter than 6 days. We study the effects of survey characteristics 
such as cadence, season, and campaign length, and derive requirements for time 
delay cosmology: in order not to bias the cosmology determination by $0.5\sigma$, 
the mean time delay fit accuracy must be better than 0.2\%. 
\end{abstract}

\date{\today} 

\maketitle

\section{Introduction} 

Strong lensing time delay cosmography is a promising probe that has developed 
rapidly in the last few years. In 2012, two strong lens time delay 
distances, combined with then-current cosmic microwave background (CMB) data, 
demonstrated as much constraining power on dark energy density and spatial 
curvature as then-current baryon acoustic oscillation distance data 
\cite{suyu12}. In 2013, a single time delay distance combined with CMB data 
determined the Hubble constant to 7\% in a dark energy model (wCDM), while 
CMB data alone nearly filled its prior \cite{suyu13}. A Hubble Space Telescope 
program to more than double the number of precisely modeled time delay lens 
systems is now underway \cite{holicow}. 

The two main cosmological advantages that strong lensing time delay distances 
bring are: 1) sensitivity to the Hubble constant $H_0$, since the time 
delay distance is a dimensionful quantity, measured from an observable time 
delay, and 2) excellent complementarity with other probes when constraining dark 
energy properties such as time varying equation of state \cite{lin04,lin11}. 
On top of this, the time delay distance is a geometric quantity, independent 
of the details of the growth of structure or galaxy bias. For further discussion 
of time delay distances as a 
cosmological probe, see \cite{snowsl}. 

Currently, the main observational challenges for the use of time delay distances 
are finding 
a large sample of lensed systems, photometrically monitoring them every 
few days over a period of several years, and following them up spectroscopically 
to establish redshifts and with high resolution imaging to model the lens 
galaxy mass distribution. The first issues will become moot with the current 
and next generation of wide field, time domain surveys such as Dark Energy 
Survey (DES; \cite{des}) and the Large Synoptic Survey Telescope (LSST; 
\cite{lsst}). Likewise spectroscopic redshifts can be obtained efficiently 
with new multiobject spectrographs such as DESI \cite{desi} and PFS \cite{pfs}. 
High resolution imaging may become easier as the HST time becomes less 
oversubscribed, the more powerful JWST is in operation, and ground based 
adaptive optics develops further. 

The major analysis challenges are the robust estimation of the actual time 
delays between images, derived from noisy, gappy lightcurves, and the modeling 
of the lens mass distribution and the mass along the line of sight. We 
concentrate on the first of these, and indeed it is the focus of a series of 
Strong Lens Data Challenges \cite{tdc0,tdc1}. The mass modeling is also developing 
rapidly \cite{oguri07,suyu09,suyu10,suyu12,collett,greene} and all three sources 
of uncertainty 
must be reduced together to obtain time delay distances with 5\% or better 
precision and subpercent accuracy. 

In Sec.~\ref{sec:gp} we describe our application of the Gaussian Process 
statistical technique to time delay estimation. We review the Time Delay 
Challenge metrics in Sec.~\ref{sec:tdc} and present our 
original blinded analysis. Section~\ref{sec:improve} describes improvements 
to the statistical methodology and their results. We discuss cosmological 
requirements on accuracy to obtain next generation constraints on the Hubble 
constant and dark energy in Sec.~\ref{sec:cos} and conclude in 
Sec.~\ref{sec:concl}.

\section{Gaussian Process Technique} \label{sec:gp} 

We employ Gaussian Process (GP) regression to estimate the time delays between 
the multiple image lightcurves of a strongly lensed source. GP is commonly used as a robust and fairly model-independent technique for reconstructing an underlying function from noisy measurements. In GP regression, the underlying function is not parametrized but instead a complete set of possible curves is fitted to the data points. The curves are constructed from a mean function, describing the average behavior of the function, and a \textit{covariance kernel} imposing a Gaussian correlation between the data points and serving to describe the fluctuation of those points around the mean function. The  covariance function is characterized by a set of \textit{hyperparameters} which control the amplitude and length of the correlation between the data points. 

Our data here are the lightcurve magnitude measurements of multiple images of 
a source. (We should emphasize that here our focus is extracting accurate time 
delays, not modeling the intrinsic lightcurve of the source.) While the measurements are made on the same underlying intrinsic lightcurve, there is a time delay between each pair of the observed lightcurves, and that is what we want to 
determine. We quantify the time delays with a set of $\Delta t_i$ parameters and fit them to data together with the GP hyperparameters of our kernel function. 
As described in detail in \cite{GPpaper}, the kernel function includes different terms:  the GP kernel (as described above) that describes the intrinsic variability of the source (generally a quasar);  a separate \textit{microlensing} kernel that accounts for the 
(longer term) variations in magnitude due to microlensing (from substructure 
in the lensing galaxy and along the line of sight); and a \textit{nugget} term, an additional constant variance in measurements that acts as a zero lag dispersion accounting for e.g.\ misestimated measurement noise or scatter due to the finite realization nature of the data.

To fit the parameters, we utilize the GP likelihood
\cite{gpml}:
\begin{equation} 
2 \ln \mathcal{L}(Y|\vec\theta)= - Y^T K^{-1} Y  - \ln |K| - N_d\, \ln 2\pi ,
\label{GP-likelihood} 
\end{equation} 
where $Y$ is the vector of magnitude data, with $N_d$ the total number of 
data points, $\vec\theta$ represents the fit parameters, e.g.\ time delays, 
and $|K|$ is the determinant of the kernel $K$, giving a complexity penalty. 

For the mean function, we adopt a constant value in this analysis as a good choice that preserves the distribution of data points, and hence any distinct features in the intrinsic source lightcurve, a necessary element in recovering accurate time delay by matching the observed lightcurves. 
We have tested other mean functions, including smoothing techniques, and found 
they did not perform as well. 

For the covariance, we adopt two kernels, a damped random walk (DRW), which is 
often used to model the intrinsic quasar/active galactic nucleus (AGN) 
light curve \cite{drw1,drw2,drw4}, 
\begin{equation} 
k(t_i,t_j)=\sigma^2\,e^{-|t_i-t_j|/l} \ , 
\end{equation} 
and a Matern function with index $3/2$ commonly used in statistics 
\cite{gpml}:
\begin{equation}
k(t_i,t_j)=\sigma^2 \left(1+\frac{|t_i-t_j|\sqrt{3}}{l}\right)\, 
e^{-|t_i-t_j|\sqrt{3}/l} \ . 
\end{equation}
In the above $t_i$ and $t_j$ are measurement times, the hyperparameter 
$\sigma$ adjusts the amplitude of the kernel and $l$ functions as a correlation length.

We use Minuit \cite{minuit}  as the likelihood minimizer, and also independently validate our fits from a Monte Carlo analysis.  The use of two kernels, two 
optimizers, and variations of priors allow us to crosscheck our results and 
determine their robustness. The GP code is parallel and optimized to handle 
a large number of systems autonomously 
in an efficient manner.

\section{Time Delay Challenge -- Blind Results} \label{sec:tdc} 

The Strong Lens Time Delay Challenge (TDC) \cite{tdc0, tdc1} provided the opportunity for the first systematic study of the current capabilities of the community in measuring time delays through a set of several thousand simulated lightcurves. The goal has been to evaluate whether the available methods were able to achieve the criteria required for handling next generation data and provide a diagnostic tool for improvements, and also to investigate the impact of different observational and systematic factors on the results.

For the TDC simulated data \cite{tdcweb}, 
an ``Evil Team'' generated LSST-like lightcurves, including 
noise and systematics, without revealing the process or true time delay, and 
released the blinded data. The intrinsic AGN light curves were constructed 
from a DRW stochastic process and then different observational, photometric  and systematic noise components were implemented progressively \cite{tdc0,tdc1}. First, microlensing contributions were added based on a simulated star magnification map for LSST. The dominant statistical noise contribution, sky brightness, was then included through a Gaussian random noise.  On top of that, additional flux errors 
were implemented in the form of three types of ``evilness'' contaminating some 
of the simulated systems. 

The main challenge (TDC1) consists of five rungs to cover a range of different observational strategies, namely, monitoring cadence and its dispersion, individual 
season length, and full campaign length. The details are summarized in Table I of \cite{tdc1}.

The TDC proposed in advance the following criteria (metrics) to evaluate the performance of methods:

\begin{itemize}

\item Submitted fraction, $f$:
\begin{equation}
\label{eq:fraction}
f \equiv \frac{N_{\rm sub}}{N} \,
\end{equation}
the fraction of the total number of systems $N$ for which time delays 
were estimated. 
\item Goodness of fit:
\begin{equation}
\label{eq:chi2}
\chi^2 = \frac{1}{fN} \sum_i \left( \frac{\tfit_i-\ttrue_i}{\sigma_i} \, \right)^2
\end{equation}
where $\ttrue_i$ is the true time delay value for system $i$, and $\tfit_i$ 
and $\sigma_i$ are the estimated time delay and its uncertainty. 

\item Accuracy (or bias):
\begin{equation}
\label{eq:accuracy}
A = \frac{1}{fN} \sum_i  \frac{ \tfit_i-\ttrue_i}{|\ttrue_i|}\ .
\end{equation}
This metric is important for getting an unbiased estimation of 
time delay distances and propagates directly into accurate determination of 
cosmological parameters. We discuss the cosmological requirements in 
Sec.~\ref{sec:cos}; TDC1 had a goal of $A<0.2\%$ \cite{tdc0} for next generation 
surveys. 

\item Precision:
\begin{equation}
P = \frac{1}{fN} \sum_i  \frac{ \sigma_i}{|\Delta t_i|} \,
\end{equation}
which quantifies the fractional uncertainty in the time delays. 

\end{itemize}

To estimate the time delays, we first run our GP code on the TDC data and fit the model parameters using both kernels and both optimizers. We then pass or reject 
each system, based on the consistency of fits and their likelihoods, and then assign a final best fit time delay and uncertainty. Finally, we rank our systems and give them confidence classes based on a set of selection criteria, a combination 
of the degree of consistency of estimated time delays from different kernels/minimizers, likelihoods, and reduced $\chi^2$. 

For TDC1, we produced six different samples, with the main three representing progressively inclusive fit confidence, e.g.\ gold, silver, bronze: \textit{Lannister}, \textit{Targaryen}, and \textit{Baratheon}. In addition, we studied other selection criteria: a  conservatively selected sample (\textit{Tully}) and one with tighter error assignment (\textit{Stark}). We also developed an outlier detection algorithm to identify and remove catastrophic outliers  through imposing controlled priors, and also an analysis of the best fit parameters for the selected systems. The \textit{Freefolk} sample was
the result of such analysis. 
The details of the statistics for these samples can be found in \cite{tdc1}. 

Our effort has been mainly focused on two aspects: developing an automated method that can handle the large number 
($\sim5000$) of future strong lens systems, fast and efficient with minimum human labor requirements; and optimizing for the accuracy of the fits as a critical 
metric for using strong lensing time delays as an unbiased, robust cosmological 
distance probe. 

After the deadline for the submission of TDC results, we revisited our code 
to study alternate mean functions and realized that our original step 
to ``prewhiten'' the lightcurves had not been fully implemented as intended. 
We made this correction, keeping everything else the same, so now the magnitude 
shift hyperparameter only has to account for residual shifts. 
This resulted in a significant improvement in the performance of our method. 
For example, for our leading blind submission of \textit{Stark} the 
average fit success fraction for the harder rungs 1-4 climbed from  $f\approx0.18$ 
to $f\approx0.33$. 

In the next section, we demonstrate that our results are accurate well below TDC requirements for the $A$ metric, and with reasonable precision ($P$), fraction ($f$) and goodness of fit ($\chi^2$).

\section{Improvements to Time Delay Estimation} \label{sec:improve}

\subsection{Criteria} 

To this point, we have followed the code output blindly, and used the 
TDC framework criterion of $\chi_i^2<10$ to cut significant outliers. 
However, $\chi^2$ requires knowledge of the true time delay and so is 
not suitable for actual cosmological use. Therefore we now add some 
intelligence to the code, while maintaining uniform treatment for all 
systems. 

The first condition considered is basic, and indeed could have been applied 
from the beginning if we had not wanted to test the fitting code in its 
purest form. 

\begin{itemize} 
\item Fit uncertainty: If the fit cannot deliver an uncertainty smaller than 
4 days, i.e.\ $\sigma(\tfit)<4$ days, then remove the system. 
\end{itemize} 

The second condition involves the accuracy. This is intended to clip 
extreme outliers. Since we will not know the true time delay, we cannot use 
it directly. However, we can identify outliers, not from the unknown true 
cosmology but from the cosmology derived from the global fit of all the 
time delay distances. That is, we compare a system against its peers. 
This statistical technique is frequently used in astrophysics, for example 
with supernova distance \cite{union}. We take a very loose clipping, 
corresponding roughly to $4\sigma$: 

\begin{itemize} 
\item Global outlier: If the fit deviates by more than 20\% from the truth, 
i.e.\ $|\tfit-\ttrue|/|\ttrue| >0.2$, then remove the system. 
\end{itemize} 
Since the global accuracy is good, and TDC1 provides no redshift information 
to derive a cosmology, we here take the global fit cosmology (and hence $\Delta t$ 
in this expression) to be the truth. 

We emphasize that the crucial uncertainty $\sigma$ here is not that of the time delay 
estimation but of the entire time delay distance estimation, i.e.\ the
cosmology estimation, including the 
uncertainties from other effects such as lens mass modeling and line of sight 
convergence. We take that the final uncertainty for the time delay distance 
of a system used 
for cosmology will be of order 5\%; thus a 20\% deviation in the time delay 
(translating into a 20\% deviation in the distance, aside from contributions and 
covariances from the other quantities) will clearly stand out. 

Those are the only two conditions we impose on our fits. We do not cut in 
$\chi^2$ or for time delays shorter than 10 days.

\subsection{Baseline results} 

Now we can examine the statistics for our improved set of fits, using 
the correct mean function treatment and the two conditions. 
Table~\ref{tab:newce} summarizes the evaluation metrics by rung.

\begin{table}[!htb]
\begin{tabular}{l|cccc} 
Rung\ \ & $\qquad f\qquad$ & $\qquad \chi^2\qquad$ & $\qquad P\qquad$ & $\qquad A\qquad$ \\ 
\hline 
   0 &  0.48 &  1.07 &  0.0578 & -0.0005 \\
   1 &  0.36 &  1.11 &  0.0617 & -0.0010 \\
   2 &  0.31 &  1.14 &  0.0854 & -0.0000 \\ 
   3 &  0.29 &  1.67 &  0.0688 & -0.0019 \\
   4 &  0.36 &  1.92 &  0.0909 & -0.0036 \\ 
\hline 
 Avg &  0.36 &  1.36 &  0.0717 & -0.0014 \\ 
 Avg[3d] & 0.36 & 1.22 & 0.0669 & -0.0008 \\ 
\end{tabular} 
\caption{Time delay estimation metrics are given for each rung of the 
challenge, and averaged over either all systems used or all systems with 
mean 3 day cadence (rungs 0-3). 
} 
\label{tab:newce} 
\end{table}

On average about one-third of the systems can be used robustly for time delay 
cosmology. Given that LSST will find of order $10^{3-4}$ systems \cite{ogmar} 
and we will be 
limited by followup observationally and by modeling uncertainties theoretically, 
such a fraction is quite acceptable. The fits achieve a mean accuracy, i.e.\ 
the bias with respect to the true time delay, of 
0.14\%; we address the cosmology requirements for this in the next section. 
The mean statistical precision is 7.2\% and is seen to be improved by more 
data in the lightcurve, either a longer season (rung 0) or longer monitoring 
campaign (rung 1). It can also be reduced by the square root of the number of 
systems. We return to the precision in Sec.~\ref{sec:vary}. 

Apart from the effect of the number of lightcurve points, the major effect is 
that the six day cadence of rung 4 performs noticeably worse than the three 
day cadence rungs. The last row of Table~\ref{tab:newce}, with only the three 
day cadence rungs, shows that the mean accuracy metric improves by almost a 
factor two, and that the precision for rung 4 is also significantly worse. 
Fixing the average cadence to three days, we see that rung 3 (with a $3\pm1$ day 
cadence) has some advantage over rung 2 (fixed 3 day cadence), as its cadence 
variation allows some sampling on shorter 
time scales; we discuss this further below. 

\begin{figure}[htbp!]
\includegraphics[width=\columnwidth]{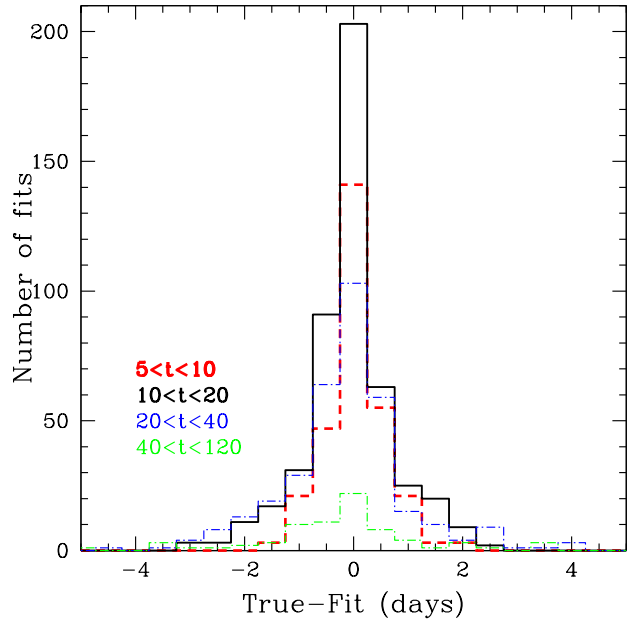} 
\caption{The distribution of the difference between the fitted time delay and 
the truth is plotted for four ranges of true time delay $t\equiv|\ttrue|$. No 
bias is apparent, and the distributions are well peaked. 
} 
\label{fig:histdttime} 
\end{figure}

It is useful to 
look at the full distributions to find more subtle effects. First we consider 
whether there is any bias in estimation for time delays of different lengths. 
Figure~\ref{fig:histdttime} plots the histograms of the deviation of fit from 
truth for four ranges of time delays. The distributions are well peaked around 
zero and fairly symmetric. The longest time delays have the broadest 
distribution but fractionally are comparable, i.e.\ a 2 day offset in a 
60 day time delay is like a 0.5 day offset in a 15 day time delay.

\begin{figure}[htbp!]
\includegraphics[width=\columnwidth]{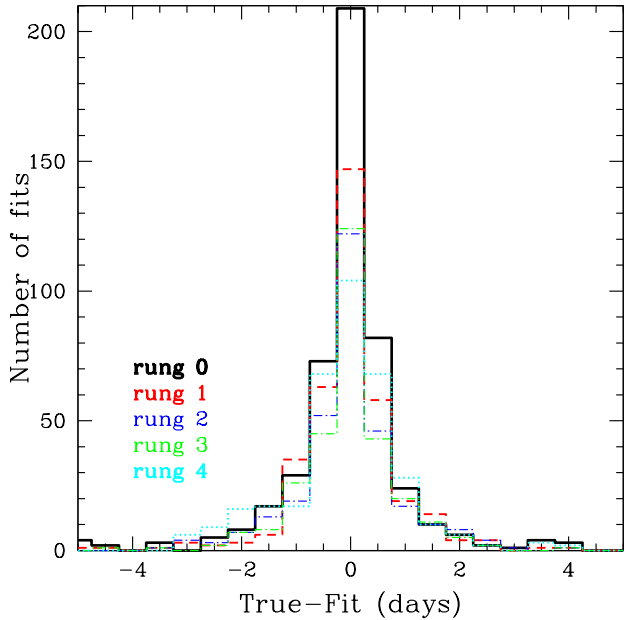} 
\caption{The distribution of the difference between the fitted time delay and 
the truth is plotted for the five sets of survey characteristics corresponding 
to the Challenge rungs. No bias is apparent, and the distributions are well 
peaked, though the result of rung 4 with six day cadence is noticeably broader. 
} 
\label{fig:histdtrung} 
\end{figure}

To study the effect of cadence and other survey characteristics, we investigate 
the distributions of results for different rungs of the challenge. 
Figure~\ref{fig:histdtrung} demonstrates that for all rungs the time delay 
estimation has negligible bias and is highly peaked around zero deviation from 
the truth. For all rungs except rung 4, the ratio of the peak to the shoulders, 
i.e.\ the height of the zero bin vs the next bins, is $\sim2.5$; however 
rung 4 with twice as long an average cadence gives a ratio of $\sim1.5$, being 
more dispersed though still unbiased. This indicates that loosening the cadence 
from three days to six could impact the cosmology results. Over rungs 0-3, 
the fit offset is less than 0.5 (1.0) days for $\sim62\%$ ($\sim82\%$) of the 
systems; for rung 4 the numbers are 52\% and 75\% respectively.

\begin{figure}[htbp!]
\includegraphics[width=\columnwidth]{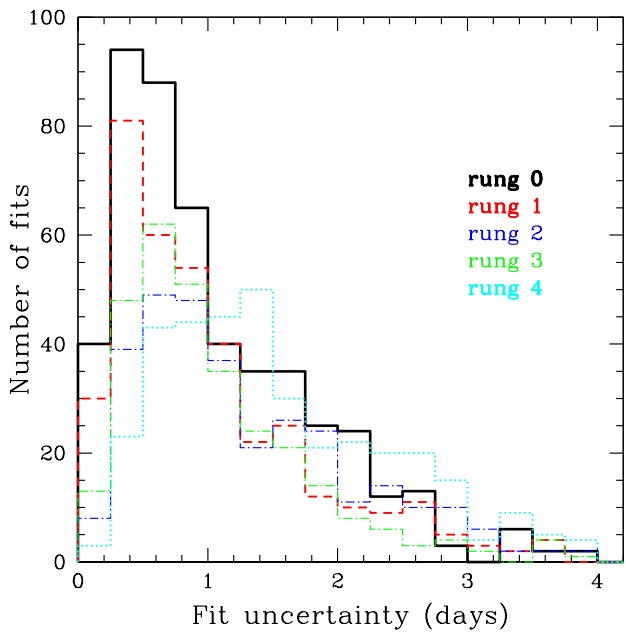} 
\caption{The distribution of the time delay fit uncertainty is plotted for the 
five sets of survey characteristics corresponding to the Challenge rungs. The 
lowest uncertainty is seen for those cases with more lightcurve points. A 
longer average cadence is deleterious, while a somewhat smaller advantage comes 
from having occasional rapid cadence observations for fixed average cadence. 
} 
\label{fig:histerrrung} 
\end{figure}

While our main focus is on accurate fits, we can also examine the impact of 
survey characteristics on statistical uncertainty of the fits. 
Figure~\ref{fig:histerrrung} shows the distributions for the various rungs. 
The number of data points play a larger role here: rungs 0 and 1, with twice 
as many lightcurve points, have smaller uncertainties. There is also some 
difference between rungs 2 and 3, where rung 2 keeps a strict three day 
cadence while rung 3 varies it between two and four days. Rung 3 has a tighter 
distribution of fit uncertainties, hinting that such variation can be 
advantageous, with the occasional tighter cadence presumably allowing 
better crosscorrelations between the images at some points in the monitoring. 
Rung 4, with the six day cadence, has a distribution of fit uncertainties 
that is noticeably shifted to longer values. While rungs 0 and 1 have fit 
uncertainties less than 0.5 (1.0) days for $\sim30\%$ ($\sim60\%$) of the 
systems, rung 4 has them for only 7\% and 32\% of the systems. Rung 3 has an 
advantage over rung 2, with 21\% vs 15\% (59\% vs 47\%) fit to better than 
0.5 (1.0) days.

\subsection{Short Time Delays} \label{sec:short} 

Note that short time delays, while 
difficult to measure precisely, can be useful. Short delays arise from either 
small time delay distance (low redshift) or small difference in Fermat 
potential, with the latter due to either very symmetric image configuration 
or small image separation. Low redshift lenses are crucial for Hubble constant 
determination; \cite{lin11} found that the estimation of $H_0$ degrades by 
55\% without $z_l<0.3$ lens systems (while higher redshift lenses are more 
useful for the dark energy equation of state; also see the systematics study in 
Sec.~\ref{sec:cos}). They are easier to follow up and model as well, with little 
line of sight mass convergence. Symmetric images can be useful as well and are 
similarly good for modeling systematics. Small image separations, however, are 
more difficult to follow up due to the limited number of pixels for the 
modeling and possibly blending of the quasar and lens light in the spectroscopy. 
Future data challenges including image information will be useful in 
investigating short delay systems in more detail.

\subsection{Variations} \label{sec:vary} 

The accuracy metric shows excellent results, with bias at only the 0.1\% 
level. We can explore some variations in the conditions to see whether 
the precision can be improved. For example, if we impose the auxiliary 
condition that $\sigma(\tfit)/|\tfit|<0.15$ (note this is not the precision 
since we use the fit $\tfit$, not the truth, and so this is a blind selection), 
i.e.\ removing fits that are 
not well constrained, then the average precision becomes 5.6\%, with the 
average fraction of systems fit reduced to 0.325. Using 
$\sigma(\tfit)/|\tfit|<0.1$ improves the precision further to 4.5\%, with 
the fraction decreasing to 0.28. In current work we have focused on obtaining 
unbiased results; future work will address improvements in uncertainty 
estimation. 

Recall that in \cite{lin11} only 150 lens systems were used to project 
cosmological constraints, and this had strong leverage. If we use only 150 
systems in a given rung, choosing those with lowest $\sigma(\tfit)/|\tfit|$, 
then we obtain precisions ranging from 1.6\% (rung 0) to 3.5\% (rung 4). 
The average accuracy over all rungs is $-0.11\%$, and over the four rungs with 
three day mean cadence is $-0.02\%$. 

\begin{figure}[htbp!]
\includegraphics[width=\columnwidth]{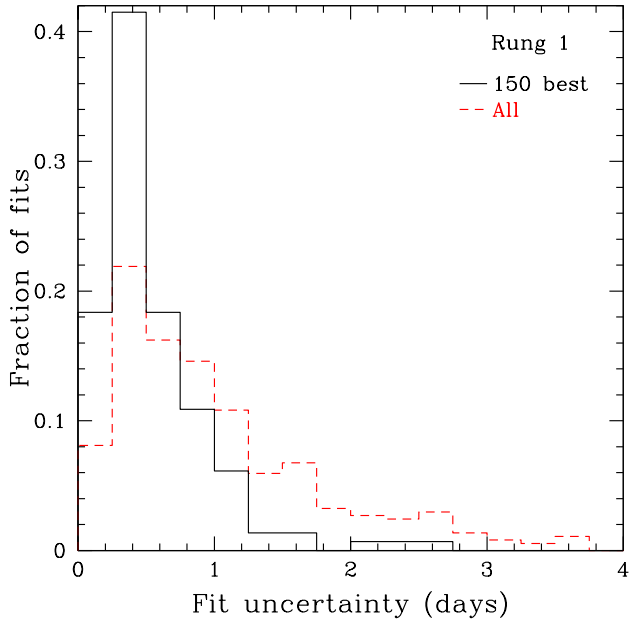} 
\caption{The distribution of the time delay fit uncertainty is plotted for the 
150 time delay estimations with lowest $\sigma(\tfit)/|\tfit|$ (solid) 
compared to all (dashed), for rung 1. The set of 150 (which may be sufficient 
for cosmology leverage) has a significantly more precise distribution. 
} 
\label{fig:histerr150} 
\end{figure}

Figure~\ref{fig:histerr150} shows the improvement in the fit uncertainty. 
Now rung 1 has 60\% (89\%) of fits within 0.5 (1.0) days, using the 150 systems 
with 
lowest $\sigma(\tfit)/|\tfit|$, compared to the previous 30\% (61\%) for all 
systems in the rung. For rung 3 the numbers are 37\% (83\%), compared to the 
previous 21\% (59\%). Table~\ref{tab:best150} summarizes the statistics for 
the time delay estimation averaged over the rungs.

\begin{table}[!htb]
\begin{tabular}{l|cc} 
Average & $\qquad P\qquad$ & $\qquad A\qquad$ \\ 
\hline 
 All rungs\ & 0.027 & -0.0011 \\ 
 3 day cadence\ & 0.025 & -0.0002\\ 
\end{tabular} 
\caption{Time delay estimation statistics are presented for the 150 time 
delays with lowest $\sigma(\tfit)/|\tfit|$ in each rung, averaged over either all rungs 
or all rungs with mean 3 day cadence (rungs 0-3). 
} 
\label{tab:best150} 
\end{table}

\section{Cosmological Requirements on Accuracy} \label{sec:cos} 

In this section we aim to quantify requirements on the accuracy of the 
time delay estimation in order for the time delay distance to be a robust 
cosmological probe. Requirements on precision can be traded off 
against more systems, since this is a statistical uncertainty, but an 
actual bias in the time delay, and hence time delay distance, can mislead 
our cosmological conclusions. 

We adopt the combination of cosmological probes used in \cite{lin11}: a 
strong lensing survey giving 1\% distance measurements in each of six 
lens redshift bins from $z_l=0.1$--0.6, together with a midrange supernova 
survey out to $z\approx1$ and Planck-quality CMB information on the distance 
to last scattering and the physical matter density $\omhh$. Such a combination 
was calculated in \cite{lin11} to deliver estimation of $\Om$ to within 0.0044, 
the reduced Hubble constant $h$ to 0.0051, or 0.7\%, and the dark energy equation 
of state today $w_0$ to 0.077 and its time variation $w_a$ to 0.26. 

A bias in the time delay $\Delta t$ leads to a bias in the time delay distance 
$D_{\Delta t}$ of the same fractional magnitude. If there were no redshift 
variation of the bias, and the only cosmological constraint came from strong 
lensing alone, then this would show up purely as an offset $\delta h$ in the 
derived Hubble constant, of the same fractional magnitude since the Hubble 
constant sets the distance scale. If we wanted a 1\% accurate Hubble constant 
measurement from strong lensing, we would need to ensure that the time delay 
bias $A$ was under 0.01. However, in the presence of other cosmological 
information, from supernovae and CMB, this simple mapping no longer holds. 
Moreover, the bias $A$ may well be redshift dependent. 

The current Time Delay Challenge does not yet incorporate cosmological 
information in the supplied lightcurves, i.e.\ there are no lens or source 
redshifts or image geometries assigned. This is planned for a future challenge. 
However, we might expect that higher redshift lens systems suffer from increased 
photometric noise and microlensing, which would affect the time delay 
estimation, as well as lens modeling (e.g.\ velocity dispersion measurement) 
and line of sight mass uncertainties. Therefore we take a phenomenological model 
of the bias 
\beq 
A(z)=A_0\,\left(\frac{1+z_s}{2.05}\right)^n \ . 
\eeq 
where $A_0$ is the amplitude, normalized to the midrange of the source 
redshift $z_s$ distribution, and $n$ allows us to vary the redshift dependence of 
the bias. Recall we took bins of lens redshift from $z_l=0.1$--0.6, 
and we assume for simplicity $z_s=3z_l$ (see \cite{lin11} for further 
discussion). 

To propagate the offset in time delay, and hence time delay distance, to 
the bias on the cosmological parameters we employ the standard Fisher bias 
formalism \cite{knox98,linbias}. The parameter bias is 
\beq 
\delta p_i=(F^{-1}){}_{ij} \sum_z \frac{\partial\tdd}{\partial p_j}\, 
\frac{1}{\sigma^2(\tdd)}\,\Delta\tdd \ , 
\eeq 
where $F$ is the Fisher matrix (here for the combined probes) 
and for simplicity we take a diagonal noise 
matrix. Note that $A=\Delta\tdd/\tdd$. The parameter bias will scale 
linearly with the amplitude $A_0$. 

\begin{figure}[htbp!]
\includegraphics[width=\columnwidth]{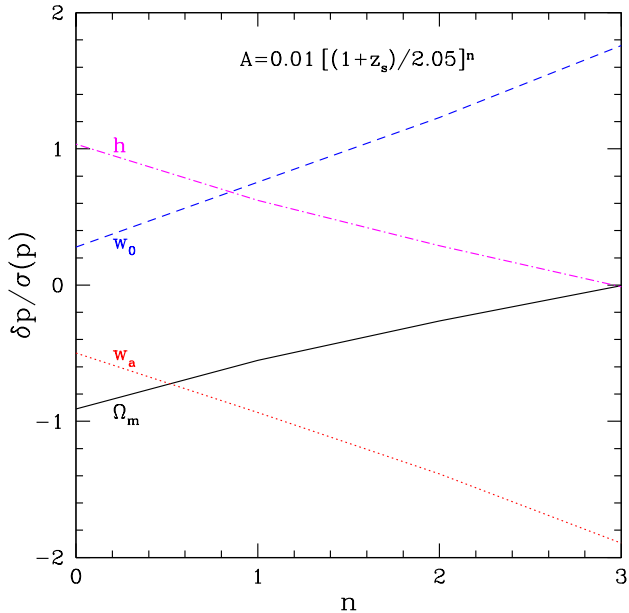} 
\caption{Bias in time delay estimation propagates into cosmological parameter 
bias, with amplitude depending on the size of the misestimation (here $A_0=0.01$ 
and we use the combination of probes mentioned in the text) 
and its redshift dependence, here taken as having power law index $n$. The 
parameter bias $\delta p$ as a fraction of the parameter uncertainty $\sigma(p)$ 
is plotted vs $n$ for the various cosmological parameters. 
} 
\label{fig:dparsig} 
\end{figure}

Figure~\ref{fig:dparsig} plots the cosmology bias of an inaccuracy 
with $A_0=0.01$ for various redshift dependences $n$. Note the nearly 
equal and opposite behavior of $\Om$ and $h$, and $w_0$ and $w_a$, due to 
their strong covariances. A redshift independent bias ($n=0$) indeed mostly 
affects the Hubble constant (and $\Om$ from its covariance), while one that 
increases rapidly with redshift predominantly affects $w_a$, since it requires 
a high redshift lever arm to see the dark energy equation of state time 
dependence.

\begin{figure}[htbp!]
\includegraphics[width=\columnwidth]{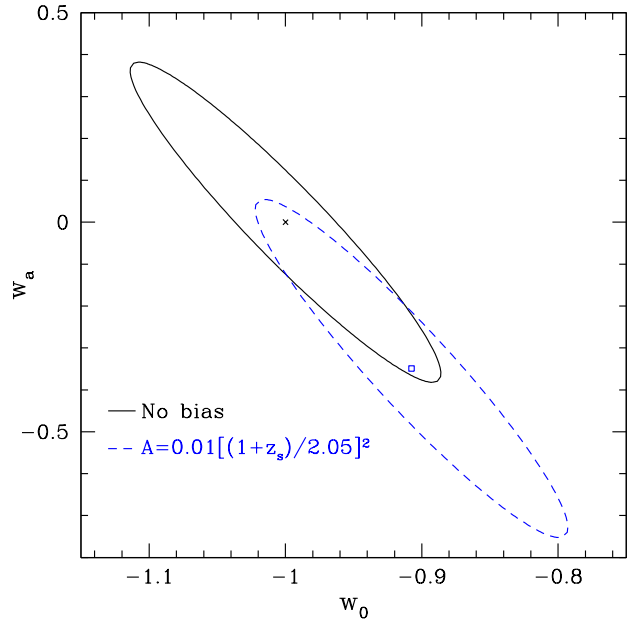} 
\caption{The 68\% joint confidence contour for the dark energy parameters 
$w_0$ and $w_a$ gets shifted by a 1\% time delay estimation bias such that 
the true cosmology (cosmological constant, marked by x) is near the edge of 
the contour. The biased value (marked by the square) falsely implies a time 
varying dark energy. 
} 
\label{fig:ellip} 
\end{figure}

Figure~\ref{fig:ellip} visualizes the cosmology bias caused by such a 1\% 
bias in time delay estimation, for the case of $n=2$.  The dark energy equation 
of state parameters are misestimated such that the derived joint values barely 
lie within the $1\sigma$ joint confidence contour of the true model, or 
conversely the true model barely lies within the derived $1\sigma$ joint 
confidence contour. To avoid such incorrect cosmological conclusions, the 
time delay must be fit more accurately.

To impose a time delay accuracy systematic requirement based on controlling 
cosmological bias, we need to specify in which parameter we are interested 
and what is the redshift dependence of the systematic. The latter is unknown 
(though future data challenges may inform this). For example, if the fit 
bias is proportional to the inverse signal to noise, then this goes as 
inverse square root of the image flux, or as the luminosity distance. Over 
the redshift range of interest, in a universe close to $\Lambda$CDM the 
angular diameter distance is roughly constant with redshift, and so the 
luminosity distance goes as $(1+z)^2$. Thus one might guess $n=2$ is roughly 
reasonable. We will also be interested in all the cosmological parameters, 
not just the Hubble constant, say, so we use Fig.~\ref{fig:dparsig} in a 
rule of thumb sense to say that a bias amplitude $A_0=0.01$, over a reasonable 
range of $n$, leads to a roughly $1\sigma$ parameter shift on some cosmological 
parameter. 

We would like the bias to be a small fraction of the statistical uncertainty 
of the cosmological parameter, $\sigma(p)$. In the presence of both statistical 
uncertainty and bias, one can use the statistical quantity of the risk, 
\beq 
R=\sqrt{(\delta p)^2+\sigma^2}=\sigma\,\sqrt{1+(\delta p/\sigma)^2} \ . 
\eeq 
We might ask that the risk increase the error over the statistical 
contribution by no more than 20\%, corresponding to $\delta p/\sigma<0.66$. 
Since $A_0=0.01$ gave $\delta p/\sigma\approx1$, then this implies we want 
$A_0<0.0066$. 

The time delay estimation is not the only contribution to the systematic 
error budget, however; there is also lens modeling, line of sight mass 
convergence, etc.\ so we adopt that the time delay bias -- being the most 
accessible to control -- should be less than 1/3 of the total systematic 
$A_0$. Putting this all together we find the requirement that 
\beqa 
A_{\Delta t} & \lesssim & \frac{0.01}{3}\, 
\frac{(\delta p/\sigma)_{\rm desired}}{(\delta p/\sigma)_{A_0=0.01}}\\ 
& \lesssim & 0.0022 \ . 
\eeqa 
We see from Table~\ref{tab:newce} that our GP time delay estimation method 
can satisfy this requirement, except in the case of the six day cadence 
(rung 4).

\section{Conclusions} \label{sec:concl} 

The time delay distance from strong gravitational lensing multiple images 
provides a unique, dimensional probe of cosmology. It is directly sensitive 
to the Hubble constant and has strong complementarity with other probes in 
determining dark energy characteristics. With new generations of surveys, 
hundreds to thousands of time delay systems will be found. We investigated one 
of the leading current sources of uncertainty for this cosmological 
probe: time delay estimation from noisy, gappy lightcurve data. 

Using a Gaussian Process statistical technique we have demonstrated control 
of systematic bias at the 0.1\% level (with precisions at 2.7\% for a 
cosmologically useful data set). The analysis was originally carried out 
on the blind mock data of the Strong Lens Time Delay Challenge. We have implemented 
an efficient, completely automated pipeline for fitting thousands of lightcurve 
systems, with delays from 5-100 days. 

The Time Delay Challenge provided data sets with different combinations 
of mean cadence, cadence variation, season length, and campaign length, allowing 
us to study their influence on the fits. We find that the number of data points 
is the most significant influence for delays from 5-100 days, but this can come 
from either longer seasons or campaigns of more years. For the rare delays of 
100 days or more, sufficiently long seasons are crucial. Lengthening the mean 
cadence raises the systematic bias, with the average three day cadence 
delivering 0.08\% accuracy but a six day cadence degrading this to 0.36\%. 
For a fixed mean cadence, sampling that allows some shorter time monitoring 
improves the precision. 

We investigated the cosmology parameter bias induced by systematic time delay 
misestimation. Depending on the redshift dependence of the systematic, the 
major effect is either on the Hubble constant or dark energy equation of state. 
As a rule of thumb, a 1\% total systematic amplitude gives a $1\sigma$ shift 
in the cosmology. Taking into account the other error contributions this 
implies that the time delay accuracy requirement should be at the 0.2\% level 
so as not to significantly bias cosmology. The GP fitting technique has 
demonstrated results sufficient to pass this requirement. 

Further improvements are under study. For example, seasons could be weighted 
by noise to remove periods of bad weather or particularly egregious 
microlensing. Our GP method delivers the full distributions of hyperparameters, 
and these could be used to study both the intrinsic variability of the 
quasar source and the microlensing. The only data provided in the challenge 
were the lightcurves; future studies could fold in image characteristics and 
other lens system information.

\acknowledgments 

We thank Alex Kim, Arman Shafieloo, Sherry Suyu, and the Evil and Good Teams 
of the Time Delay Data Challenges for useful discussions (and the Evil Team 
for exemplary effort in generating the Challenges), the Institute 
for the Early Universe, Korea for computational resources, and IBS and KASI for 
hospitality. 
The simulated lightcurve data used in this work was generated by the Strong 
Lens Time Delay Challenge ``Evil Team'' (Liao, Dobler, Fassnacht, Marshall, 
Rumbaugh, Treu) and is available at \url{http://timedelaychallenge.org}. 
EL was supported in part by NASA, DOE grant DE-SC-0007867 and 
the Director, Office of Science, Office of High Energy Physics, of the 
U.S.\ Department of Energy under Contract No.\ DE-AC02-05CH11231.  AH is 
supported by an NSERC grant.


\end{document}